\documentclass[prl,twocolumn,aps,showpacs]{revtex4}

\begin{document}

\title{Comment on: Detecting Vanishing Dimensions
Via Primordial Gravitational Wave Astronomy}

\author{Thomas P. Sotiriou$^{1}$, Matt Visser$^{2}$, and Silke Weinfurtner$^{3}$}

\affiliation{$^1$DAMTP,
Centre for Mathematical Sciences,
University of Cambridge, Wilberforce Road, Cambridge CB3 0WA, UK\\
$^2$School of Mathematics, Statistics, and Operations Research,
Victoria University of Wellington, PO Box 600, Wellington 6140, New Zealand\\
$^3$SISSA - ISAS,
Via Bonomea 265, 34136, Trieste, Italy
{\rm and} INFN, Sezione di Trieste}
\date{\today}

\maketitle

It has been argued~\cite{detecting} that quantum gravity models where the number of dimensions reduces at the ultraviolet (UV) exhibit a potentially observable cutoff in the primordial gravitational wave (GW) spectrum. It was claimed there that this is a ``generic" and ``robust'' test for such models,  since ``$(2+1)$-dimensional spacetimes have no gravitational degrees of freedom".
We find this claim misleading for two distinct reasons:

{\em Definition of dimensionality:} It is ambiguous as to which definition of ``dimension" is being used when referring to vanishing dimensions~\cite{detecting}. The only papers cited there which discuss vanishing dimensions in quantum gravity are~\cite{CDT,shirkov}, which discuss the  ``spectral dimension'' (SD). 
In particular~\cite{CDT}  refers to casual dynamical triangulations [CDTs], where it is the SD that
  flows to 2 in the UV, {\em not} the topological [physical] dimension. The latter remains 4~\cite{CDT}. It is not true that CDTs ``demonstrate that the four-dimensional spacetime can emerge from two-dimensional simplicial complexes'', as stated in~\cite{detecting}. 4-dimensional CDTs by construction arise from 4-dimensional simplices.  The SD is an analytic feature that provides information about short-distance dispersion relations. The fact that it runs to $2$ at short distances does not mean that any physical modes decouple there.

Consider, for instance, Ho\v{r}ava gravity, which is another theory where (Lorentz violating) dispersion relations lead to a SD of 2 in the UV \cite{Horava}.  The spin-2 graviton does not decouple at high energies, it remains part of the physical excitation spectrum, albeit with a strongly Lorentz violating dispersion relation. 

That is, in a wide class of models where some notion of dimension is scale dependent, this is the SD. But the SD is not the quantity that appears in the Feynman loop integrals (as in the suggestions made in~\cite{detecting}); that is the physical dimension, 4, which is not running.

{\em Dimensionality and dynamics:} Though there are some heuristic models cited in~\cite{detecting} where it is the physical dimension that is running, {\em e.g.}~\cite{lattice}, these are quantum field theory models which do not include gravity. 
Let us nevertheless entertain the idea that  it is indeed the number of physical dimensions that reduces in the UV in a quantum gravity model and one ends up with a lower-dimensional theory. The argument used  in support of the claim that such a theory would have no local degrees of freedom is essentially that $2+1$ dimensional general relativity (GR) has this property~\cite{detecting}. However, there is no particular reason to believe that a generic quantum gravity model which reduces to a $2+1$ dimensional theory at high energies should share this characteristic.

$2+1$ dimensional gravity theories can most certainly have local degrees of freedom. GR is dynamically trivial in $3$ dimensions due to the specific structure of its field equations. $3$ dimensional spacetimes can  be dynamical, and so would be the theory that would describe them. Note in particular that, even though the Weyl tensor vanishes identically in 3 dimensions, this does not imply that the metric is conformally flat. This would require that the Cotton tensor vanish, which is not generically true. Clearly, a $3$ dimensional theory cannot have gravitons with the usual polarizations, simply because there are not enough spatial dimensions, but this does not mean that it will not have local degrees of freedom at all.

Additionally, there is no reason whatsoever for the theory in question to be close to $2+1$ dimensional GR in the UV. Clearly, if this is a to be a viable gravity theory it should resemble 4-dimensional GR at low energies. In fact, it is difficult to imagine how a theory that has this latter property can  have the former property as well.

{\em Conclusions:} Vanishing dimensions in quantum gravity are certainly expected to leave an imprint on primordial GWs, but this imprint need by no means be a ``generic" and ``robust'' absence of any gravitational radiation in the sense of a sharp cutoff,
as claimed in~\cite{detecting}. 
Such models can have local degrees of freedom in the UV, as it can either be that it is some other notion of dimension that drops below $4$, not the physical dimension, or that they simply differ from 3 dimensional GR.

\vskip - 10 pt


\end{document}